# Orientations of LASCO Halo CMEs and Their Connection to the Flux Rope Structure of Interplanetary CMEs

V. Yurchyshyn[a], Q.Hu[b], R.P. Lepping[c], B.J. Lynch[d], and J. Krall[e]

[a]Big Bear Solar Observatory, 40386 North Shore Lane, Big Bear City, CA 92314, USA
[b]IGPP, University of California, Riverside, CA 92521, USA
[c]NASA, Goddard Space Flight Center, Code 696.0, Greenbelt, MD 20771, USA
[d]SSL, University of California, Berkeley, 7 Gauss Way, Berkeley, CA 94720
[e]PPL, Naval Research Laboratory, Code 6790, Washington, DC 20375-5000 USA

**Abstract**

Coronal mass ejections (CMEs) observed near the Sun via LASCO coronographic imaging are the most important solar drivers of geomagnetic storms. ICMEs, their interplanetary, near-Earth counterparts, can be detected in-situ, for example, by the Wind and ACE spacecraft. An ICME usually exhibits a complex structure that very often includes a magnetic cloud (MC). They can be commonly modelled as magnetic flux ropes and there is observational evidence to expect that the orientation of a halo CME elongation corresponds to the orientation of the flux rope. In this study, we compare orientations of elongated CME halos and the corresponding MCs, measured by Wind and ACE spacecraft. We characterize the MC structures by using the Grad-Shafranov reconstruction technique and three MC fitting methods to obtain their axis directions. The CME tilt angles and MC fitted axis angles were compared without taking into account handedness of the underlying flux rope field and the polarity of its axial field. We report that for about 64% of CME-MC events, we found a good correspondence between the orientation angles implying that for the majority of interplanetary ejecta their orientations do not change significantly (less than 45 deg rotation) while travelling from the Sun to the near Earth environment.

## 1. Introduction

Earth-directed coronal mass ejections (CMEs) are often seen in the LASCO coronagraph on board SOHO as bright expanding halos surrounding the Sun (Howard et al., 1982). CMEs are believed to be the result of a large-scale rearrangement of the solar magnetic field (Low, 2001, Schwenn et al. 2006) and, when observed near the Earth, their magnetic structure, or parts of it, can variously be described as complex ejecta (Burlaga et al., 2001), magnetic clouds (MC, Burlaga et al., 1981, Bothmer and Schwenn, 1998), plasmoids or shocks, associated with twisted IMF with foot points rooted in the Sun (Dryer, 1994). Generally, MCs exhibit a magnetically organized geometry, which is thought to correspond globally to a curved flux rope (Bothmer and Schwenn, 1998).

LASCO images provide us with the 2D projection of a 3D structure, and therefore the projection effect may greatly affect the observed shape and speed of CMEs. Halos often exhibit various sizes and shapes. Many of them can be enveloped by an ellipse and fitted with a cone model (Zhao et al., 2002, Xie et al., 2004, Zhao, 2005, Michalek et al., 2006).

CMEs that erupt close to the solar limb and are seen from the side (Figure 1) often exhibit a three-part structure with the bright leading edge around the dark cavity and a bright core at the bottom (Dere et al., 1999). This structure is commonly interpreted as an expanding magnetic flux rope (Chen et al., 1997, Low, 2001). Cremades and Bothmer (2004) examined 124 limb events and concluded that their white-light morphology bears information of the magnetic structure. These authors also argue that the structured CMEs are magnetically organized in the axial direction, which corresponds to the axis of a large-scale twisted flux rope. The flux rope like features are not uncommon in CMEs (see Figure 1; Fig. 2 of Li et al. (2001); Fig. 5 in Manoharan et al. (2001); Fig. 1 in Krall et al. (2006); also Krall 2006). A recent study by Krall & St Cyr (2006) showed that the statistical parameters such as eccentricity and the axial aspect ratio, obtained for a parameterized flux-rope CME is in agreement with the corresponding observed measures (St Cyr et al., 2004).

Krall et al. (2006) modelled the well-known 2003 October 28 event as an erupting flux rope and generated a synthetic coronagraph image of this model "halo" which appeared to be elongated in the direction of the flux rope axial field. Comparison of the model best fit to the observed LASCO image showed that the tilt angle of the elongation in both modelled and observed halos was similar, thus indicating that the ellipse-shaped appearance of halo CMEs may be related to their magnetic structure (see Figures 2 and 8 in Krall et al., 2006). This suggestion was further supported in Yurchyshyn et al. (2006) and illustrated in Figure 1 in this paper. Krall et al. (2006) also concluded that the flux rope's axis in this event rotated smoothly as the flux-rope apex expanded from the solar surface to 1 AU. This rotation seems to occur in the near-Sun region.

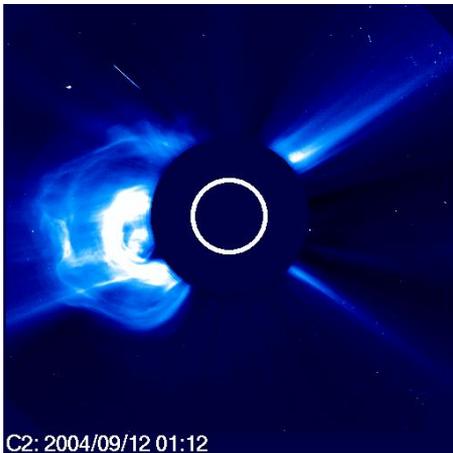

*Fig.—1. LASCO C2 image of a limb CME that erupted on 2004 September 12. This image represents and event with a distinct three part structure and suggests that, at least some CMEs, can be described as a flux rope structures.*

The findings mentioned above motivated us to study whether there is any correlation between the estimates of orientation angles of MCs and the tilt of elongated CMEs, and to understand if the elongation indicates the axial direction of an underlying flux rope. In the present paper, we describe a statistical comparison of orientations of 25 CME--MC pairs. We note that the halo and MC angles are compared without addressing handedness of the underlying flux rope field.

## 2. Data Analysis

The present study includes 25 events (Table I) selected according to the following criteria. For each CME—ICME pair we must be able to reliably identify the solar surface event (to exclude back side halo CMEs), the corresponding CME in the solar corona and the ICME at 1AU. The majority of the selected CME-ICME events are those listed in the Master Data Table compiled during a *Living With a Star* Coordinated Data Analysis Workshop (CDAW[1]). We also used the list of events published in Qiu and Yurchyshyn (2005).

The CME orientation angles were determined by fitting an ellipse to an irregularly shaped "halo" around the C3 occulting disk. Eight points, evenly spaced in position angle, were measured along the outer edge of the halo in each image available for a given event (lines in Figure 2). At each angle, the edge of the halo is chosen to be the outermost point on the overall expanding CME structure. The measured points were then fitted with an ellipse and its tilt angle, $\alpha_{CME}$, was measured in the clockwise direction from the positive $y$-axis. Because for many events the measured tilt angles, $\alpha^i_{CME}$, are scattered around some mean value, the final orientation angle, $\alpha_{CME}$, listed in Table I, was calculated as the mean of all angles, $\alpha^i_{CME}$, determined from individual images (on average, we had 3-5 LASCO C3 frames per event). Note that in Figure 2 we present the case when a halo CME did not propagate directly toward the Earth along the Sun-Earth line but was rather deflected toward the southwest. Nevertheless, even for this event, the projection of an expanding flux loop on the plane of the sky appears to have an elliptical shape and its elongation corresponds to the axial direction of the underlying flux rope.

[1] http://cdaw.gsfc.nasa.gov/geomag_cdaw/Data.html

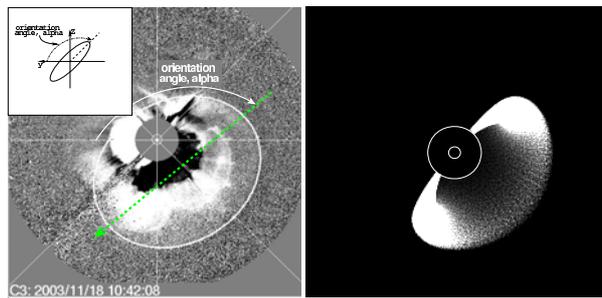

*Fig.—2. Left: LASCO C3 difference image for the 2003 November 18 eruption used to measure the CME elongation. The halo CME was measured at its outer edge at intervals of 45 degrees (lines). An ellipse was then fitted to the eight halo points and its tilt angle, $\alpha_{CME}$, was determined in the clockwise direction from the y-axis that points eastward to the ellipse semi-major axis (see inset). Right: A synthetic "halo" CME image produced for the same event by the erupting flux rope model. The tilt of the synthetic elongated halo represents the orientation of the simulated flux rope. Comparison with the left panel shows that the tilts of those two halos are similar thus indicating the correspondence between the observed halo elongation and the magnetic axis of a flux rope.*

For each event in Table I we applied a Grad-Shafranov reconstruction routine (Hu and Sonnerup 2002, herein GS) to determine the orientation (clock) angle of a MC, $\alpha_{GS}$. This angle is the direction angle of the projected MC flux rope axis onto the GSE $yz$-plane, measured in the clockwise direction from the positive $y$-axis of the GSE coordinate system ($y$-axis is in the ecliptic plane pointing towards dusk, $x$-axis directed from the Earth towards the Sun and $z$-axis is pointed upward). Note, that we used the same coordinate system to calculate all orientation angles discussed in this study. Some MCs could not be resolved by the GS routine. Successful events are listed in the fifth column of Table I.

In order to make the study more reliable and, at the same time, increase the number of events, we used the orientation angles independently determined from a MC fitting routine (Lepping et al., 1990) by Lynch et al. (2005, herein LY), $\alpha_{LY}$, and by the Wind MFI Team (Lepping et al., 2006, herein MFI), $\alpha_{MFI}$ (Columns 6 and 7 in Table I). Note, that the original fitting parameters were given in terms of longitude and latitude in GSE coordinates, so they were converted in clock angles in the GSE coordinate system. For three events (2003 October 28, 2003 November 18 and 2005 May 13) we used orientation angles produced by the erupting flux rope (EFR) model (see Krall et al. (2006) for details of MC fitting). Please note that for some events the orientation angles are clustered near 360-degree mark: in these cases, to ease the comparison, the 360 deg were added to those orientation angles that are within the 0 ... 90 deg range.

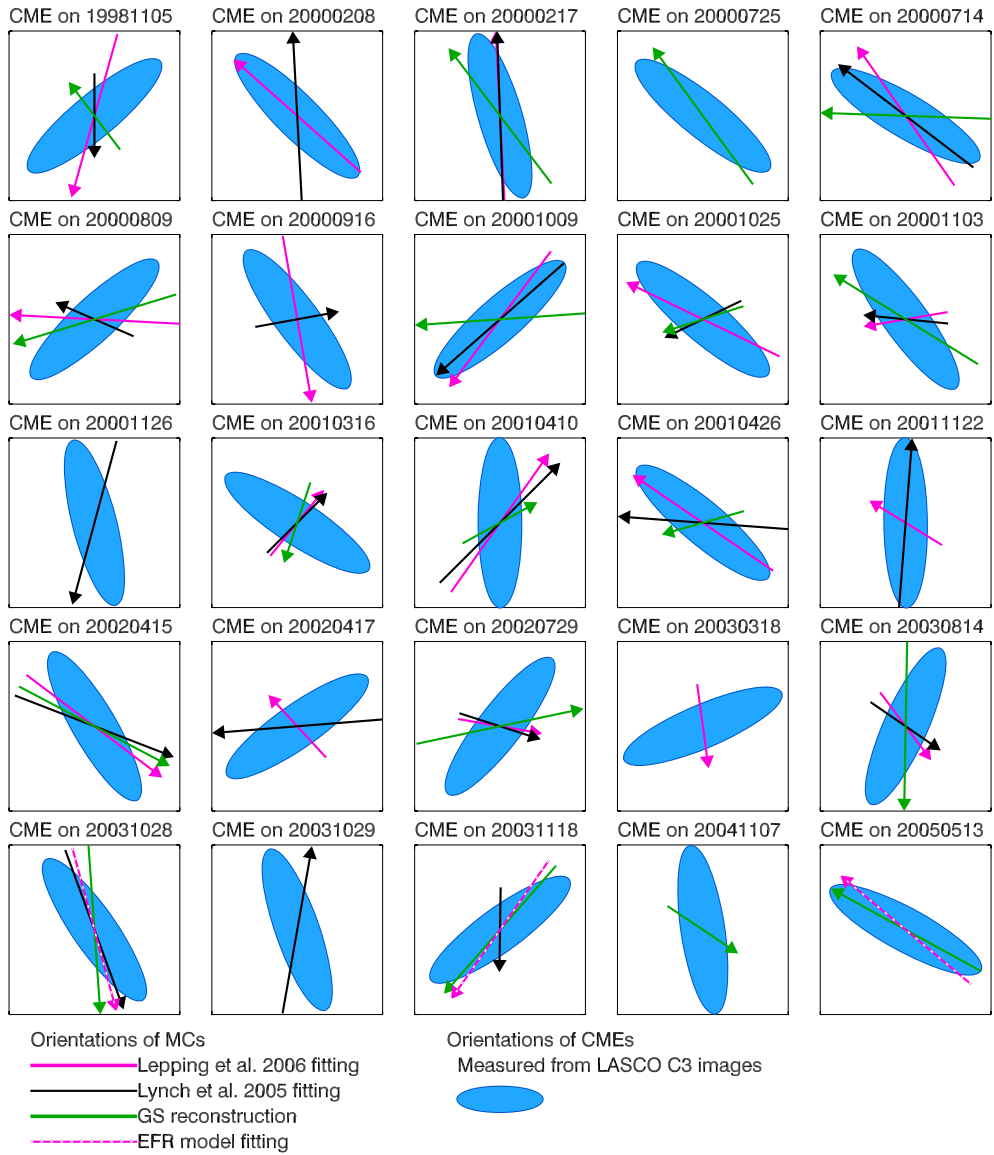

Fig.—3. Comparison between the LASCO CME (filled ellipses) and MC clock angles (arrows). Black arrows represent the Lynch et al. (2005) fitting, red arrows the Lepping et al. (2006) fitting, green arrows the GS reconstruction and dotted red arrows the EFR model. Long (short) arrows indicate cases when the difference between the LASCO and MC clock angles is smaller (exceeds) than the threshold of 45 deg.

## 3. Results

Figure 2 shows comparison of the CME and MC clock angles. The major axes of the filled ellipses indicate the orientation of observed LASCO CMEs, while various arrows represent results from the GS reconstruction (green), MC fitting by Lynch et al. (2005, black) and Lepping et al. (2006, red) and the EFR model (dotted red). To estimate the correspondence between the LASCO and MC orientations we calculated the difference, $\Delta\alpha$, between the CME and MC angles. We then determined the ratio of the CME — MC pairs with Δα < 45 deg to the total number of events (correspondence rate). Those MCs, whose clock angles differ from the CME orientation(s) by more than 45 deg, are shown in Figure 2 with short arrows. We should note that this comparison was made by applying an acute angle method when the difference is taken as a minimum angle between the line and the semi-major axis without considering the polarity of the MC axial field. Two methods had nearly the same tightly clustered correspondence rate: GS routine -- 67% (12 out of 18 events) and MFI fitting — 61% (11 out of 18 events).

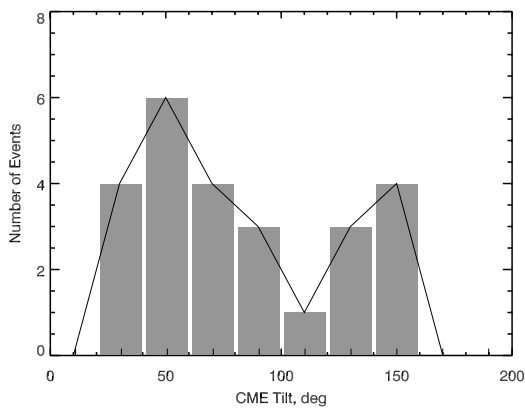

*Fig.—4. Distribution of the orientation angles of LASCO CMEs listed in the 2$^{nd}$ column of Table I. The distribution is not uniform: there are two well pronounces peaks centered at 50 and 150 degrees, which correspond to orientations on the Sun along the NE-SW and SE-NW lines. The bin size in this distribution is equal to 20 degrees.*

Lynch et al. (2005) fitting had a correspondence rate of 57% (12 out of 21 events). We would like to point out, however, that in two cases (1998 November 5 and 2001 April 10) LY results were similar to the corresponding MFI angles yet, they did exceed the threshold by several degrees. If we consider these two events then LY success rate will approach 67%. All three events analyzed by the EFR model ($α_{EFR}$ = 255, 305 and 40 deg for the 20031028, 20031118 and 20050513 events, respectively) showed good agreement between observed LASCO orientations and MC clock angles. We would like to emphasize that in the case of a random distribution of CME and MS orientation angles, one would expect approximately a 50% correspondence rate at the threshold angle of 45 deg. The observed rate exceeds this estimation. (Note that the derived correspondence rate is about 49% for the threshold of 35 deg, while the random correspondence rate is 39%.) We thus conclude that our study of 25 CME — ICME events shows that, on average, the orientation of two thirds of halo CMEs does not change significantly (less than 45 deg rotation), during their propagation in the heliosphere and interplanetary media toward the near Earth environment.

## 4. Discussion

We would like to briefly discuss the fact that only about 2/3 of the analyzed events showed a good correspondence between the LASCO halo CME tilt angle and the MC orientation angle. First, the inferred correspondence rate could be affected by shortcomings of the methodology and techniques used here, which allow us to determine the MC axis position with accuracy no better than 20 deg. Also, some of the CMEs in this data set are not full halo, therefore their orientation may not be accurately determined. Second possibility is that a CME may be an ice-cream cone with a circular base, whose projection on the plane of the sky will be an ellipse, unless this ice-cone CME is propagating strictly along the Sun-Earth line. The orientation of an ellipse, in this case, will be such that its major semi-axis will always be parallel to the solar limb, regardless of the CME propagation. Therefore, one may expect that tilts of the elongated CMEs should be randomly distributed. Figure 4 shows distribution of the orientation angles of CME elongations. Please, note that for this plot 180 or 360 deg where subtracted from all $α_{CME}$ > 180 deg (see Table I, 2$^{nd}$ column). As it follows from the bar plot, the distribution is not uniform and there are two distinct peaks centered at approximately 50 and 150 degrees. This result, in general, argues against the ice-cream cone model, although it is quite possible that some CMEs, in our data set, whose orientation does not agree with that of the corresponding MC may represent an ice-cream cone eruption.

Another feasible interpretation of these findings is that the main axis of a CME rotates as it expands into the interplanetary space. It is generally viewed that the wavy and spiral heliospheric current sheet (HCS, Schultz, 1973, Smith et al., 1978) interacts with the solar ejecta that moves through the heliosphere (Zhao and Hoeksema, 1996). Crooker et al. (1993) suggested that the base of the HCS may often include multiple helmet streamers and that most CMEs might then be spatially associated with the HCS, which can be considered as a conduit for outward propagating CMEs. Whenever the ejecta are moving at a faster speed than the upstream plasma there must be an upstream influence. About 1/2 of these events drive upstream shocks, testifying to the fact of their "superior" speed in general (Gosling et al., 1994, Howard and Tappin, 2005). Therefore, it may be expected that the heliospheric magnetic field (HMF), which is one of the upstream structures, should influence a CME's motion by draping around the ejecta, thus deflecting it from the initial direction and, quite possible, rotating the axis of the CME as it moves through interplanetary space (Smith, 2001). However, details of the dynamics of the interaction between the heliospheric structures and CMEs are not well studied yet and the scale of the effect that HMF has on CMEs, is largely unknown. Clearly, the problem of ICME evolution in the interplanetary media, which is a complex dynamic system that includes the Sun, solar wind and the magnetosphere, needs to be further studied.

**Acknowledgement**

We thank the ACE MAG and SWEPAM instrument teams and the ACE Science Center for providing the ACE data and A. Szabo and F. Mariani of the Wind/MFI team for data management and calibration. SOHO is a project of international cooperation between ESA and NASA. VY thanks V. Abramenko for discussions and useful suggestions. VY work was supported under NSF grant 0536921 and NASA ACE NNG0-4GJ51G and LWS TR&T NNG0-5GN34G grants. QH is supported by NASA NNG04GF47G. JK's work was supported by NASA (DPR W-10106, LWS TRT Program) and the Office of Naval Research.

Table 1. List of studied CMEs, their corresponding MCs and their orientation angles measured in degrees clockwise from the positive direction of the y-axis in the GSE system. Column 1 represents date and time of CMEs; column 2 is the LASCO CME orientation angle, $\alpha_{CME}$, in degrees; columns 3 and 4 are date and day of year of the corresponding MC; columns 5-7 show the orientation angle of a MC as determined from Hu and Sonnerup (2002) routine ($\alpha_{GS}$), and reported by Lynch et al. ($\alpha_{LY}$, 2005) and Lepping et al. ($\alpha_{MFI}$, 2006).

| CME Date | $\alpha_{CME}$, deg | MC Date | DoY | $\alpha_{GS}$, deg | $\alpha_{LY}$, deg | $\alpha_{MFI}$, deg |
|---|---|---|---|---|---|---|
| 19981105 20:44 | 320.1 | 19981109 | 313 | 413 | 270.0 | 285.6 |
| 20000208 09:30 | 45.2 | 20000212 | 43 | . | 87.0 | 41.8 |
| 20000217 21:30 | 73.9 | 20000221 | 52 | 53 | 88.0 | 87.1 |
| 20000725 03:30 | 39.2 | 20000728 | 210 | 54 | . | . |
| 20000714 10:54 | 30.9 | 20000715 | 197 | 2 | 37.5 | 55.2 |
| 20000809 16:30 | 383.7 | 20000812 | 225 | 343 | 383.6 | 363.0 |
| 20000916 05:18 | 233.5 | 20000918 | 262 | . | 169.6 | 260.2 |
| 20001009 23:50 | 318.5 | 20001013 | 287 | 356 | 318.5 | 306.6 |
| 20001025 08:26 | 400.7 | 20001029 | 302 | 342 | 334.1 | 386.0 |
| 20001103 18:26 | 414.1 | 20001106 | 311 | 392 | 365.7 | 350.2 |
| 20001126 17:06 | 255.7 | 20001128 | 333 | . | 285.0 | . |
| 20010316 03:50 | 213.5 | 20010319 | 078 | 288 | 134.9 | 128.9 |
| 20010410 05:30 | 89.7 | 20010412 | 102 | 151 | 134.9 | 125.1 |
| 20010426 12:30 | 400.3 | 20010428 | 119 | 344 | 364.3 | 394.5 |
| 20011122 23:30 | 89.4 | 20011124 | 328 | . | 94.4 | 31.7 |
| 20020415 03:50 | 240.0 | 20020417 | 107 | 208.0 | 201.3 | 217.3 |
| 20020417 08:26 | 291.5 | 20020420 | 110 | . | 355.4 | 407.6 |
| 20020729 12:07 | 127.3 | 20020802 | 214 | 168.0 | 197.9 | 189.9 |
| 20030318 12:30 | 336.8 | 20030320 | 79 | . | . | 262.3 |
| 20030814 20:06 | 113.9 | 20030818 | 230 | 271.0 | 214.7 | 233.3 |
| 20031028 11:30 | 235.4 | 20031029 | 302 | 266.0 | 250.0 | . |
| 20031029 20:54 | 70.9 | 20031031 | 304 | . | 99.9 | . |
| 20031118 08:50 | 323.7 | 20031120 | 324 | 311.0 | 270.9 | . |
| 20041107 16:54 | 98.4 | 20041109 | 314 | 214 | . | . |
| 20050513 17:22 | 28.7 | 20050515 | 133 | 29 | 35.0 | . |